\documentstyle[11pt,newpasp,twoside,epsf]{article}
\def\fun#1#2{\lower3.6pt\vbox{\baselineskip0pt\lineskip.9pt
  \ialign{$\mathsurround=0pt#1\hfil##\hfil$\crcr#2\crcr\sim\crcr}}}

\def\edcomment#1{\iffalse\marginpar{\raggedright\sl#1\/}\else\relax\fi}
\reversemarginpar

\begin{document}
\title{Dynamics of Black-Hole Nuclei}
 \author{D. Merritt, F. Cruz, M. Milosavljevi\'c}
\affil{Rutgers University, New Brunswick, NJ, USA}

\begin{abstract}
Space-based observations are beginning to yield detailed information 
about the stellar kinematics at the very centers of galaxies, within the 
sphere of gravitational influence of the black hole.
The structure and dynamics of these regions is probably determined
in part by the infall and coalescence of black holes during galaxy
mergers. A goal of $N$-body simulations is to reproduce the kinematics 
near the black holes as well as the relations that exist between the nuclear 
and global properties of galaxies.
However, the problem is computationally difficult due to the wide range 
of length and time scales, and no single $N$-body code can efficiently
follow the evolution from kiloparsec to sub-parsec scales.
We review existing $N$-body work on this problem and present the first,
fully self-consistent
merger simulations of galaxies containing dense stellar cusps and 
black holes.
\end{abstract}

\section{Introduction}

Supermassive black holes (BHs) are generic components of galaxies,
and they appear to be linked in fundamental ways to the dynamics of the stellar component, both on large and small scales.
An astonishingly strong correlation exists between BH mass and stellar velocity dispersion, 
$M_{\bullet}\sim\sigma^{\alpha}$, $\alpha\approx 5$ (Ferrarese \& Merritt 2000;  Gebhardt et al. 2000b).
The $M_{\bullet}-\sigma$ relation as defined by the best-determined BH masses 
is so tight that it surpasses in predictive accuracy what can be achieved from detailed dynamical modelling of high-quality stellar-kinematical data in most galaxies.
On small scales, the approximately power-law variation of stellar density with radius (Ferrarese et al. 1994) would be difficult to understand in the absence of BHs.
Low luminosity ellipticals and bulges have steep nuclear density profiles, $\rho\sim r^{-\gamma}$, $\gamma\approx 2$; 
such steep cusps form naturally in stellar systems where
the BHs grow on time scales long compared to crossing times
(Peebles 1972; Quinlan, Hernquist \& Sigurdsson 1995).
Brighter galaxies typically have weaker cusps, $\rho\sim r^{-1}$
(Merritt \& Fridman 1995; Gebhardt et al. 1996).
While no universally accepted model has yet been proposed for the origin
of the weak cusps, we note here one feature that suggests a 
link to BHs. Two galaxies, NGC 3379 and M87, have weak
cusps with well-determined structural parameters and also have
BHs with accurately determined masses. Table 1 gives for
each galaxy the ``break'' radius $r_b$ at which the central power 
law turns over to a steeper outer profile; and also the stellar mass
$M_*$ contained within $r_b$.  In both galaxies, $M_*(r_b)$ is identical
within the uncertainties with $M_{\bullet}$, the mass of the BH;
this is in spite of a factor of $\sim 30$ difference in $M_{\bullet}$ 
and $\sim 6$ in $r_b$. 

The rough equality between $M_{\bullet}$ and $M_*(r_b)$, 
and the exclusive association of weak cusps with bright galaxies, 
is consistent with the suggestion that weak cusps are relics of
galaxy mergers (Begelman et al. 1980; Ebisuzaki et al. 1991).
In this model, the two BHs from the merging galaxies fall to the center 
and heat the stars, reducing the stellar density and forming a tight binary.
The ``damage'' done by the binary, and hence the extent of the shallow
cusp, would be roughly proportional to the total BH mass.

\begin{table}
\caption{ }
\begin{tabular}{lcccc}
\tableline
Galaxy & $\gamma$ & $r_b$ (pc) & $M_{\bullet}$ ($M_{\odot}$) & $M_*(r_b)$ 
($M_{\odot}$) \\
\tableline
NGC 3379 & $1.1$\tablenotemark{(1)} & $51$\tablenotemark{(1)}  & $1.35\times 10^8$\tablenotemark{(2)} & $3.5\times 10^8$ \\
M87 & $1.26$\tablenotemark{(3)} & $315$\tablenotemark{(3)} & $3.6\times 10^9$\tablenotemark{(4)}  & $4.8\times 10^9$ \\
\tableline
\tableline
\tablenotetext{1}{Gebhardt et al. 1996}
\tablenotetext{2}{Gebhardt et al. 2000a}
\tablenotetext{3}{van der Marel 1994}
\tablenotetext{4}{Macchetto et al. 1997}
\end{tabular}
\end{table}

New instruments and observational techniques, both ground-based
(e.g. SAURON; Miller 2000) and space-based (HST; e.g. Joseph et al. 2000),
are giving us increasingly detailed glimpses of the stellar kinematics 
very near the centers of galaxies.
These data are useful both for constraining BH masses, 
but also as fossil signatures of the stellar dynamical processes that shaped
nuclei.
Interpreting these new data in the context of models for galaxy formation
will be a challenging computational task.
Stellar densities in a steep stellar cusp are $\ga 10^7 M_{\odot}$ pc$^{-3}$ 
implying dynamical times of $\la 10^4$ yr, 
and even less in the vicinity of the BH;
thus an $N$-body code that simulates both the large- and small-scale evolution of merging galaxies must be able to handle a range of $\sim 10^5$ in time scales as well as a sufficiently large number of particles to resolve the stellar cusps.
Until recently, $N$-body codes satisfying these requirements were almost non-existent, but happily that situation has now changed.
Tree codes that efficiently implement variable time steps on parallel architectures now exist (e.g. {\tt GADGET}, Springel et al. 2000); such codes are ideal for following the early stages of galaxy mergers before the BHs form a bound pair.
In the later stages, when the BH binary begins to interact ``collisionally'' with individual stars, a code like {\tt NBODY6++} (Spurzem \& Baumgardt 1999) is more appropriate: this code implements Aarseth's efficient hierarchical algorithm using block time steps on parallel computers and can treat particle-particle interactions on arbitrarily small scales.

Merger simulations will need to reproduce a number of observed regularities in the properties of the stellar cusps.
One is the tight correlation between $M_{\bullet}$ and $\sigma$ mentioned
above.
Another observational result is the dependence of cusp slope $\gamma$ on 
galaxy luminosity (Gebhardt et al. 1996): 
faint ellipticals, $|M_v|\ga 20$, have $\gamma\sim 2$ while $\gamma$ for bright ellipticals varies from $\sim 2$ to $\sim 0$.
There are at least three interesting questions here: (1) What determines the characteristic luminosity separating galaxies with strong and weak cusps? 
(2) Assuming that all galaxies begin life with strong cusps, 
how do the weak cusps form?  
(3) How are the weak cusps in bright galaxies maintained in spite of mergers with dense, low-luminosity galaxies?
Another empirical result is the rough proportionality between 
break radius $r_b$ and galaxy luminosity (Faber et al. 1997).
A direct link between cusp slope and BH mass has also been suggested 
(van der Marel 1999).

In addition to these well-established parameter relations, 
the new, high-resolution kinematical data will provide 
detailed stellar velocity fields in galactic nuclei which 
can be compared to $N$-body simulations.

In this article, we summarize the existing $N$-body work on interactions between galaxies with high central densities and BHs.
We then present preliminary results from new simulations which, for the first time, follow the evolution of BH nuclei through a realistic merger to the formation and decay of a hard BH binary.

\section{$N$-Body Studies of Mergers with Stellar Cusps and Black Holes}

In the merger of two galaxies each containing a supermassive nuclear BH,
the dynamical evolution can usefully be divided into two stages, 
before and after the formation of a hard BH binary.
Quinlan (1996) defines a ``hard'' BH binary as one in which the semimajor axis 
$a$ satisfies
\begin{equation}
a < a_h = {Gm_2\over 4\sigma^2} = 2.7 {\rm pc} \left({m_2\over 10^8M_{\odot}}\right) \left({200 {\rm km\ s}^{-1}\over\sigma}\right)^2
\end{equation}
with $m_2$ the mass of the smaller of the two BHs and $\sigma$ the stellar velocity dispersion.
Before the binary becomes hard, each BH acts as an independent point mass and its orbit decays in roughly the way described by Chandrasekhar's dynamical friction formula (except that the relevant mass is that of the BH plus any stars from the original cusp that remain bound to it).
The evolution in this regime is essentially collisionless, 
i.e., independent of the masses of the background stars.
After the formation of the hard binary, the two BHs move about the galactic center like a single particle of mass $M_{12}=m_1+m_2$ and their interaction with passing stars is dominated by three-body encounters. 
The latter almost always result in the transfer of energy from the BH binary to the stars;
as a consequence, stars are ejected from the nucleus and the binary shrinks (Hills 1992; Quinlan 1996).

Simulating the merger of two galaxies,
each containing a dense stellar cusp and a central point mass,
is computationally demanding, and essentially no such simulations 
have appeared in the literature to date.
Barnes (1999) described mergers between identical spherical galaxies with Dehnen's (1993) density law, having central power-law cusps with indices $\gamma=(1,1.5,2)$, but no BHs. He used a tree code with fixed time step.
Barnes found that the $\gamma=1$ and $\gamma=1.5$ cusps became slightly shallower following the merger but that the $\gamma=2$ cusps were nearly unchanged.
Holley-Bockelmann \& Richstone (1999) obtained similar results in simulations of unequal-mass mergers, also without BHs; 
they fixed the potential corresponding to the larger galaxy and assumed that the smaller galaxy remained spherical as its orbit decayed.
They found that the secondary remained intact as long as its initial density was higher than that of the primary, as expected in mergers between real galaxies.
Neither of these studies was able to produce weak cusps in galaxies that did not contain them initially.

\begin{figure}
\plotone{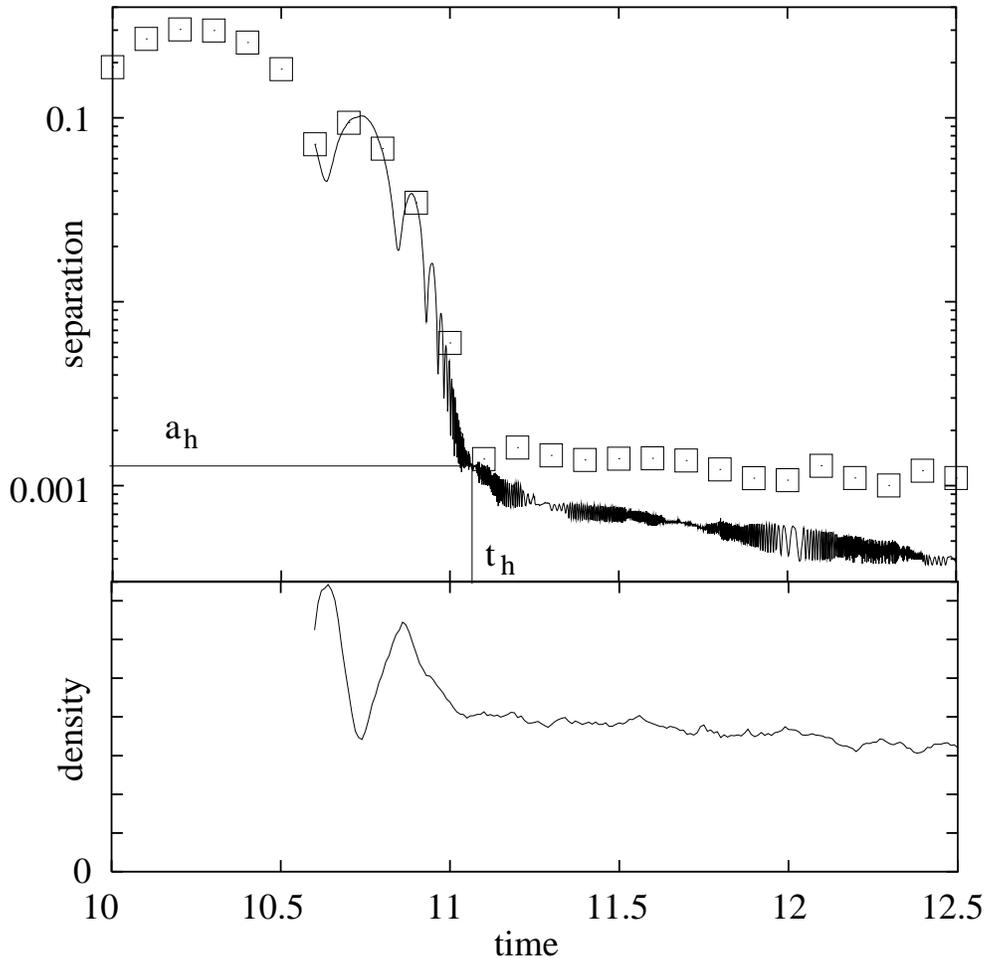}
\caption{
Upper panel: Separation between the BHs in a set of merger simulations.
Time is measured from the start of the simulation in units such that
the crossing time in a single galaxy is $\sim 2.2$.
The BHs form a hard binary at $t=t_h\approx 11.1$.
Squares are from a $1.31\times 10^5$-particle integration with 
the tree code {\tt GADGET}; 
solid line is from the {\tt NBODY6++} integration with $N=65,536$.
The binary separation saturates at roughly the softening length
$h$ in the tree code simulation, while the direct-summation 
code is able to follow the decay to arbitrarily small scales.
Lower panel: Stellar density as a function of time at the center of the 
$N=65,536$, {\tt NBODY6++} integration.
Density is defined as number of particles within a sphere of radius 
$0.05$ centered on the origin.
The density drops sharply at $t\approx t_h$ when the hard binary
forms, then decays more slowly thereafter.
}
\end{figure}

Merger simulations including nuclear BHs have been carried out but only from rather unrealistic initial conditions, typically consisting of galaxies with large, constant-density cores 
(Ebisuzaki et al. 1991; Makino et al. 1993; Governato et al. 1994; Makino \& Ebisuzaki 1996).
The BHs were found to generate still larger cores in the merger products due to heating of the stars;
the core tended to grow in such a way as to maintain rough proportionality between core radius and half-mass radius, and between core mass and (total) BH mass.
This result is consistent with observed parameter relations (Faber et al. 1997) if core radius is identified with break radius $r_b$.
The central density profiles in some of these simulations showed weak, 
power-law cusps.
Nakano \& Makino (1999a,b) argued that this was due simply to the removal of low-energy stars by heating from one or both of the BHs.
Holley-Bockelmann \& Richstone (2000) extended their earlier, BH-free simulations to the case of a fixed primary containing a central point mass and an evolving secondary with no BH.
They found that the secondary was usually disrupted by tidal forces from the primary's BH; 
however this result might be modified if a BH were added to the secondary as well.

A collisional $N$-body code is required to follow the evolution once the BH binary becomes hard.
This statement is at first sight surprising, since the rate of evolution of a hard binary in a background of low-mass perturbers depends only on their total density, not their individual masses (Hills 1992; Quinlan 1996).
However the rate of binary decay depends also on how widely the binary wanders about the center of the potential.
A fixed binary rapidly ejects all of the stars that pass near it; 
continued evolution requires that the binary wander to regions of higher density.
It is able to do this because the low stellar density implies a small gravitational restoring force,
and because the binary receives kicks from ejected stars.
The resultant random walk depends on the size of the kicks, hence on
the mass ratio $m_*/M_{\bullet}$ and the number of particles $N$.
The decay rates observed in the simulations with largest $N$, $N\approx 3\times 10^5$ (Makino 1997; Quinlan \& Hernquist 1997; Hemsendorf 1999),
imply a binary lifetime (until gravitational radiation coalescence) of order a Hubble time (Merritt 2000);
if one extrapolates the decay rate to still larger $N$ using the $N$-dependence estimated by Makino (1997), $\sim N^{-1/2}$, one concludes that BH binaries in real galaxies might be expected to often survive long enough for a third BH to find its way to the center (Valtonen 1996).
However Quinlan \& Hernquist (1997) saw evidence that the decay rate may saturate for $N\ga 10^5$.
Another possibility is that gas dynamical processes accelerate the decay (Gould \& Rix 2000).

\begin{figure}
\plotone{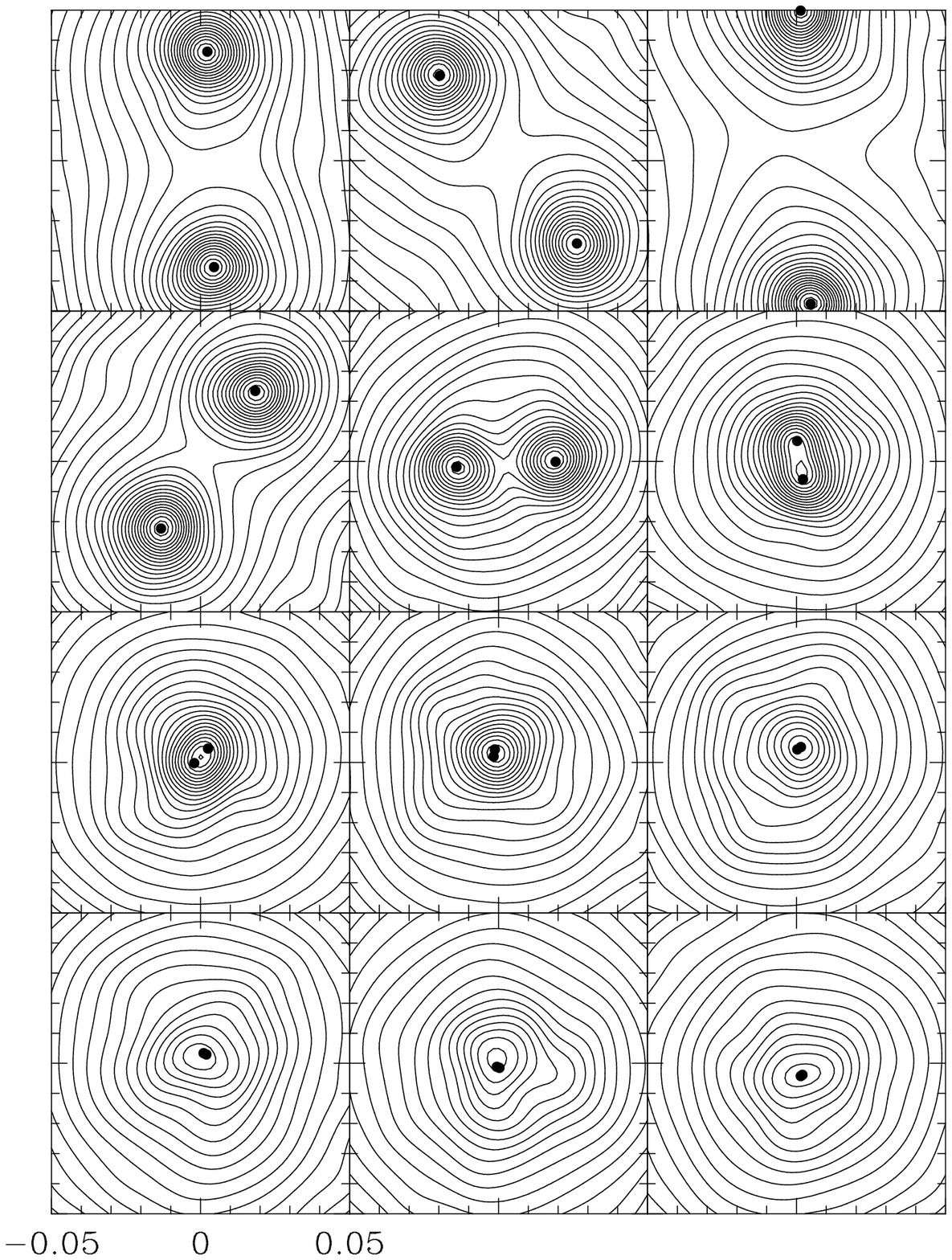}
\caption{
Transformation of two density cusps into one.
Contours show the surface density in the {\tt NBODY6++} 
merger simulation with $65,536$ particles.
Contours are separated by $0.13$ in the logarithm;
dark circles are the BHs.
Projection is onto the initial orbital plane.
Frames are separated by unequal time intervals; 
the first frame is at $t=10.6$, the last at $t=11.5$, and
the BH binary becomes hard approximately at frame $8$, 
$t=11.0$.
The two cusps remain nearly intact until after the BHs form
a hard binary, at which point the density drops suddenly
(see also Fig. 1b).
}
\end{figure}

Although the timescale for evolution of massive BH binaries to complete coalescence is still unclear, most of the damage that the binary inflicts on the stellar cusp takes place shortly after it becomes hard.
The mass ejected is 
\begin{equation}
M_{ej} \approx M_{12}\ln(a_h/a)
\end{equation}
(Quinlan 1996),
hence the binary ejects of order its own mass after shrinking by a factor 
of only two.
This is probably sufficient to convert a strong cusp into a weak cusp (Table 1).
Quinlan \& Hernquist (1997) presented density profiles that look qualitatively similar to those in bright galaxies, with a clear break radius;
some of these profiles even exhibit central minima.
The decaying binary also has a strong influence on the nuclear kinematics,
since ejected stars are predominantly on eccentric orbits; 
the remaining stars exhibit a tangentially-biassed velocity ellipsoid.
Detecting this velocity polarization in real galaxies would be difficult due to the low surface brightnesses of weak-cusp galaxies.

\begin{figure}
\plotfiddle{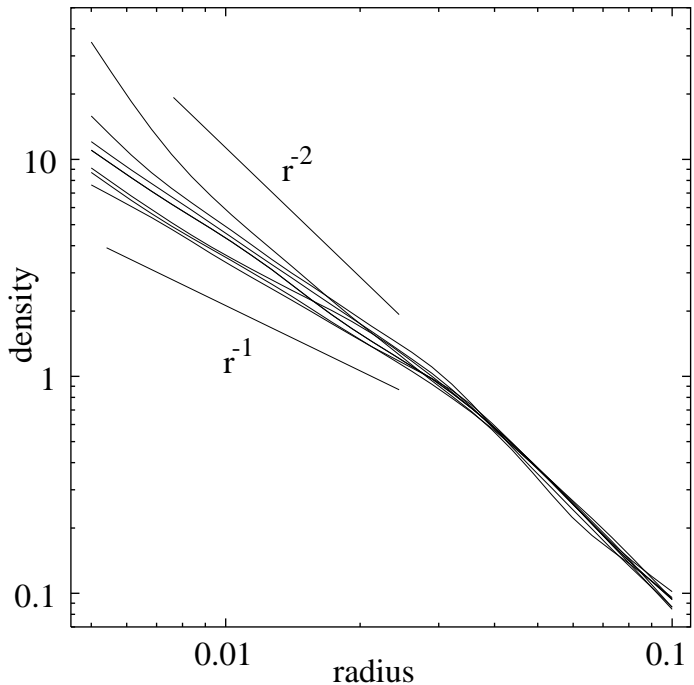}{3.75truein}{0.}{70}{70}{-140}{0}
\caption{
Stellar density profiles in the {\tt NBODY6++} merger simulation.
Top curve is at $t=11.0\approx t_h$, the time of formation of the hard
binary, and bottom curve is at $t=11.7$
The origin is defined as the center of mass of the BH binary.
The initial, $\rho\sim r^{-2}$ cusp is converted into
a shallower, $\sim r^{-1}$ cusp by the action of the binary.}
\end{figure}

\begin{figure}
\plotone{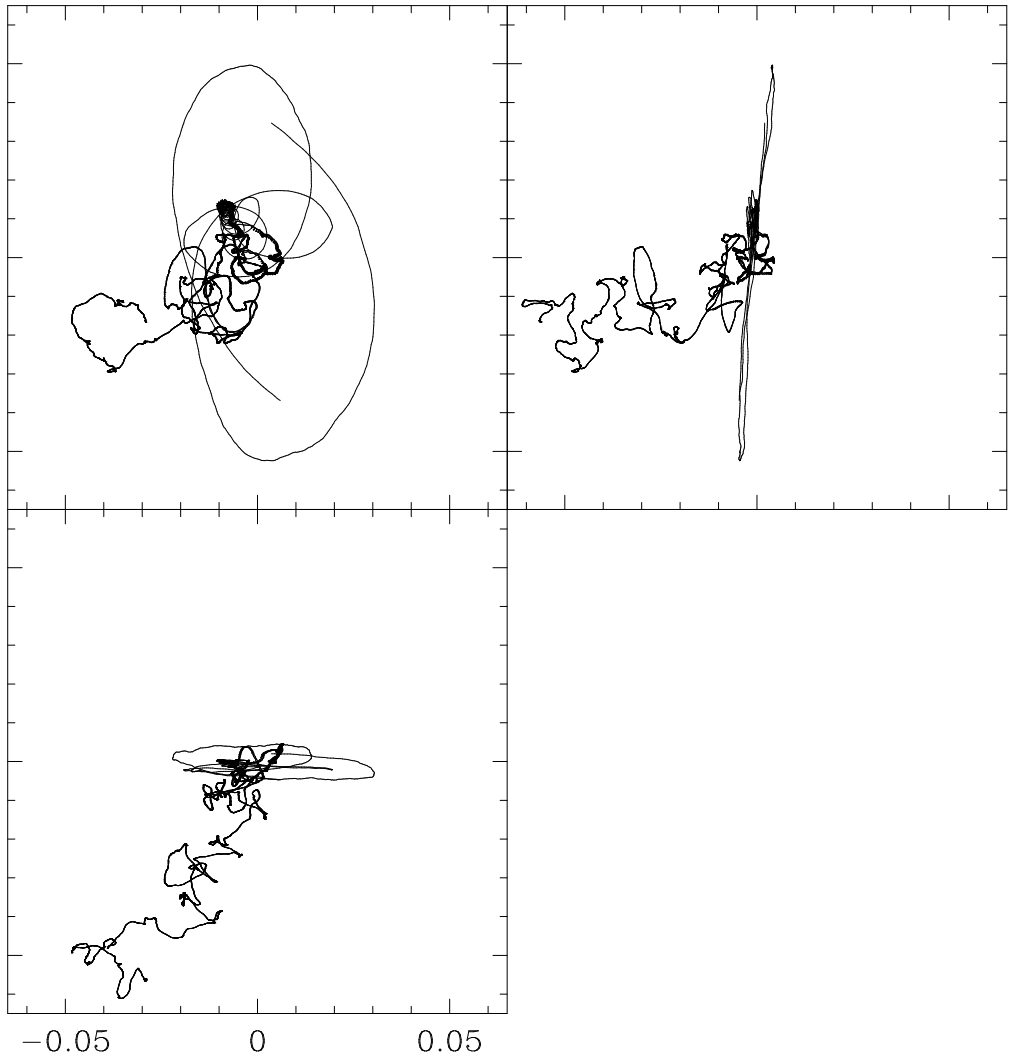}
\caption{
Wandering of the BH binary in the $32,768$-particle, 
{\tt NBODY6++} merger simulation.
Each line traces the location of one BH, from time $t=10.6$,
shortly before the binary becomes hard, to $t=20.0$.
The upper left panel is projected onto the initial orbital plane of the
two galaxies.
Wandering begins suddenly when the binary becomes hard, 
and its amplitude increases with time.
}
\end{figure}

Taken together, these simulations probably reproduce much of the interesting dynamics that takes place during the merger of two galaxies containing nuclear BHs.
Among the interesting questions yet to be answered are:
(1) Can a shallow power-law cusp be produced from the merger of two galaxies with steep cusps and nuclear BHs?
(2) Can a shallow cusp survive the accretion of a small dense galaxy with a nuclear BH?
(3) What is the precise $N$-dependence of the wandering amplitude and decay rate of a BH binary in a galactic nucleus? 
(4) Is the structure of the resultant stellar core also $N$-dependent?

In addressing these questions, one would ideally like to simulate mergers in a continuous way from the largest, kiloparsec scales down to the sub-parsec scales that characterize the BH binary.
We describe in the next section a first attempt to carry out this ambitious program.

\section{A Merger Simulation}

We report the simulation of a merger between two spherical, 
equal-mass galaxies with steep central density cusps surrounding nuclear BHs.
We first integrated the entire merger using the tree code {\tt GADGET} 
(Springel et al. 2000), a parallel algorithm with continuously 
variable time steps.
{\tt GADGET} has a fixed spatial resolution determined by a softening 
length $h$, 
which was chosen to be slightly smaller than $a_h$, 
the expected separation of the hard BH binary.
To follow the evolution of the binary at late times when $a<h$, 
we continued the integration using {\tt NBODY6++}, a parallel, 
multiple-time-step version of Aarseth's direct-summation code.
The particle coordinates as computed with {\tt GADGET} were extracted at a time before the formation of the hard binary and used to generate initial conditions for {\tt NBODY6++}.
The early evolution as computed by {\tt NBODY6++} was compared to that of {\tt GADGET} over the same time interval to ensure that the tree code had accurately integrated the equations of motion prior to extraction of the coordinates.

\begin{figure}
\plotone{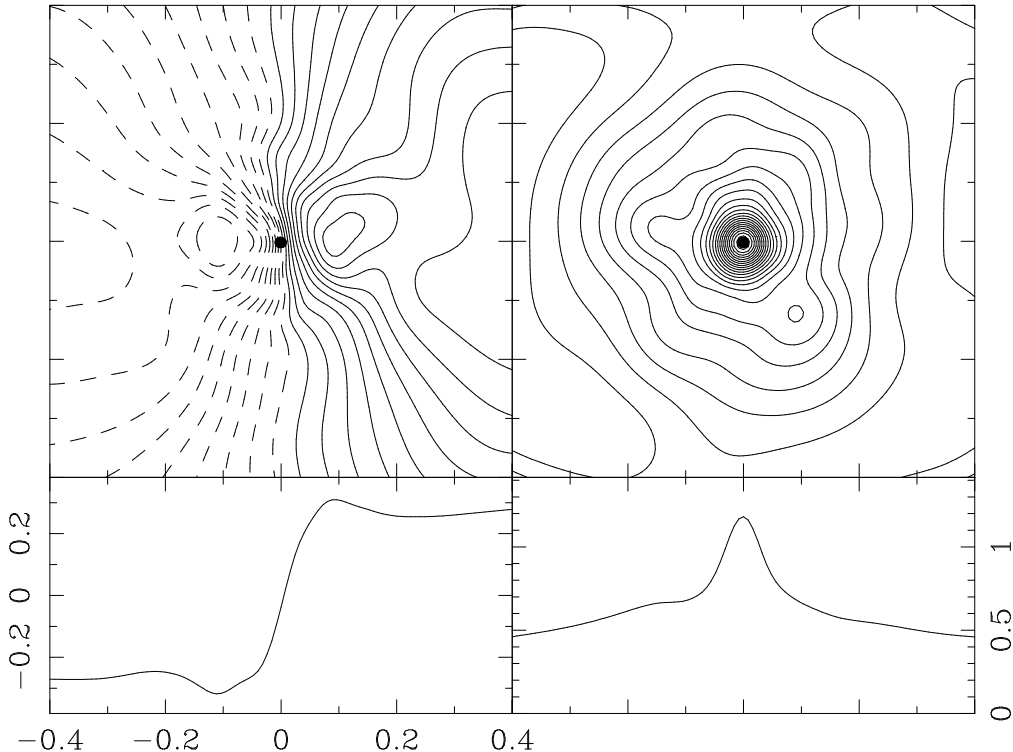}
\caption{
Stellar velocity fields at the end of the $N=65,536$ merger simulation,
as viewed from the initial orbital plane.
Left panels show contours of the mean velocity and the major-axis
rotation curve.
Right panels show the velocity dispersion.
}
\end{figure}

The initial galaxies were generated from Dehnen's (1993) law with 
central density dependence $\rho\propto r^{-\gamma}$, $\gamma=2$.
Initial velocities were assigned based on an isotropic distribution function 
that accounts for the presence of the BH (Tremaine et al. 1994).
Each galaxy was given $N=1.31\times 10^5$ equal-mass particles;
the two particles representing the BHs were given masses $1\%$  
that of their host galaxies.
Two, randomly-selected subsets of particles were extracted for the 
later integration with {\tt NBODY6++},  
with $N=32,768$ and $N=65,536$.
The initial separation between the galaxies was four times the half-mass 
radius of each, where $r_{1/2}=1.0$,
and the initial orbital velocity was $1/2$ that of a circular orbit.
The internal orbital period at the half-mass radius of either galaxy was 8.8 in $N$-body units; the {\tt GADGET} integration was carried out for $\sim 2$ of these periods and required approximately 10 days using 8 processors on the Rutgers Sun E10000 supercomputer.
At the time of this writing, the {\tt NBODY6++} integration with 
$N=32768$ had advanced until a time of $\sim 20$ 
and the integration with $N=65,536$ until a time of $\sim 12$.

The separation between the two BH particles as a function of time 
is shown in Fig. 1a.
As computed by {\tt GADGET}, the binary separation stalls at $a\approx h$,
the softening length,
soon after the BHs have fallen to the center and formed a hard binary.
With {\tt NBODY6++}, 
the binary continues to slowly decay as it exchanges energy with the 
stars.
However most of the damage done by the binary to the cusp takes place very shortly after the binary forms.
The central density (Fig. 1b) drops by a large factor in just a few orbital periods of the binary.
This behavior is consistent with a simple model (Merritt 2000) 
which predicts a mass ejected by the binary of
\begin{equation}
M_{ej} = M_{12}\ln\left[{4(t-t_h)\over t_0}\right]
\end{equation}
with $t_h$ the time of formation of the hard binary and
$t_0\approx 1.76 GM_{12}/\sigma_*^3 \approx 0.02 $ in $N$-body units.
The transformation of two density cusps into one is shown in Fig. 2;
here again, the very sudden drop in density at
$t\approx t_h$ is apparent.

Fig. 3 shows the change in the stellar density profile
during the time of its most rapid evolution, $t\ga t_h$.
The cusp evolves quickly from its initial $r^{-2}$ 
dependence to a shallower power law, $\rho\sim r^{-1}$,
and a break appears at a radius $\sim 0.03 \sim 2.4\times a_h$.
Thereafter the profile continues to evolve toward shallower slopes
but very slowly, as the binary continues to eject stars.
This figure confirms for the first time that the merger of two galaxies 
with steep cusps and BHs can produce a weak, power-law cusp.
Bright ellipticals and bulges sometimes have even shallower
central profiles (Gebhardt et al. 1996).
Increasing the masses of the BHs would probably produce weaker cusps
(e.g. Quinlan \& Hernquist 1997), although the BHs in our simulation
are already uncomfortably large.
The continued, slow evolution in the central density as the binary
ejects stars (Fig. 1b) might also produce a shallower profile;
repeated mergers would probably have have a similar effect.
Future work should be directed toward understanding which of these
explanations accounts for the observed trend of cusp slope with
galaxy luminosity.

The sharply lowered density at $t\ga t_h$ causes another change
in the behavior of the binary, as shown in Fig. 4.
The binary begins suddenly to wander over a region much larger than $a_h$.
Wandering was inferred by Quinlan \& Hernquist (1997) and Makino (1997)
in their $N$-body studies but little is known about its properties.
The wandering is enabled by the reduced force gradients
in the low-density core; it is driven by the ejection of single stars
that interact strongly with the binary and hence its amplitude
depends on the mass ratio $m_*/M_{12}$.
Fig. 4 suggests that the wandering amplitude is constant or
even increasing with time.
The $N$-dependence of the wandering amplitude is another important topic
for future study.

The kinematical structure of the merged galaxy at late times is 
shown in Fig. 5.
The rotation curve reflects primarily the initial 
orbital angular momentum of the galaxies and not the presence of the
BHs; it peaks at a radius of $\sim$ a few times $r_b$, similar to what
is seen in merger simulations without BHs (e.g. Bendo \& Barnes 2000)
and in real, high-luminosity ellipticals (e.g. Statler et al. 1999).
It is very different from the rotation curves seen in fainter
galaxies (e.g. M32, Joseph et al. 2000) which remain flat or
rising into the sphere of influence of the BH.
The velocity dispersions do continue to rise toward the center,
roughly as $r^{-1/2}$, as they did in the initial galaxy models.

\bigskip
This work was supported by NASA through grant NAG5-6037
and by the NSF through grant AST 96-17088.
We thank S. Aarseth and R. Spurzem for their advice and help with
{\tt NBODY6++}.

\end{document}